\documentclass[]{raa}            
\usepackage{graphicx,times}
\usepackage{natbib}

\providecommand{\bm}[1]{\mbox{\boldmath$#1$\unboldmath}}

\begin{document}
\title{Global energetics of solar powerful events on 6 September 2017}

\volnopage{ {\bf --} Vol.\ {\bf --} No. {\bf XX}, 000--000}
   \setcounter{page}{1}

\author{Dong~Li\inst{1,2}\thanks{Corresponding author},Alexander~Warmuth\inst{3}, Jincheng~Wang\inst{4}, Haisheng~Zhao\inst{5}, Lei~Lu\inst{1}, Qingmin~Zhang\inst{1}, Nina~Dresing\inst{6}, Rami~Vainio\inst{6}, Christian~Palmroos\inst{6}, Miikka~Paassilta\inst{6}, Annamaria~Fedeli\inst{6}, Marie~Dominique\inst{7}}

\institute{$^1$Purple Mountain Observatory, Chinese Academy of Sciences, Nanjing 210023, China; {\it lidong@pmo.ac.cn} \\
           $^2$State Key Laboratory of Space Weather, Chinese Academy of Sciences, Beijing 100190, PR China \\
           $^3$Leibniz-Institut f\"{u}r Astrophysik Potsdam (AIP), An der Sternwarte 16, 14482 Potsdam, Germany \\
           $^4$Yunnan Observatories, Chinese Academy of Sciences, Kunming Yunnan 650216, China \\
           $^5$Key Laboratory of Particle Astrophysics, Institute of High Energy Physics, CAS, Beijing 100049, China \\
           $^6$Department of Physics and Astronomy, University of Turku, Turku 20014, Finland \\
           $^7$Solar-Terrestrial Centre of Excellence/SIDC, Royal Observatory of Belgium, 3 Avenue Circulaire, B-1180 Uccle, Belgium \\
 }

\abstract{Solar flares and coronal mass ejections (CMEs) are thought
to be the most powerful events on the Sun. They can release energy
as high as $\sim$10$^{32}$~erg in tens of minutes, and could produce
solar energetic particles (SEPs) in the interplanetary space. We
explore global energy budgets of solar major eruptions on 6
September 2017, including the energy partition of a powerful solar
flare, the energy budget of the accompanied CME and SEPs. In the
wavelength range shortward of $\sim$222~nm, a
major contribution of the flare radiated energy is in the soft X-ray
(SXR)~0.1$-$7~nm domain. The flare energy radiated at wavelengths of
Ly$\alpha$ and middle ultraviolet is larger than that radiated in
the extreme ultraviolet wavelength, but it is much less than that
radiated in the SXR waveband. The total flare radiated energy could
be comparable to the thermal and nonthermal energies. The energies
carried by the major flare and its accompanied CME are roughly
equal, and they are both powered by the magnetic free energy in the
AR NOAA 12673. Moreover, the CME is efficient in accelerating SEPs,
and that the prompt component (whether it comes from the solar flare
or the CME) contributes only a negligible fraction.
}

\authorrunning{Li et al.}
\titlerunning{Global energetics of solar powerful events on 6 September 2017}
\maketitle

\section{Introduction}
Solar flares are sudden and drastic energy-release processes at
multi-height atmospheres on the Sun, which can be radiated across a
wide range of wavelengths, such as: radio, infrared, white light
(WL), ultraviolet (UV), X-rays and $\gamma$-ray beyond 1~GeV
\citep[e.g.,][references therein]{Benz17}. A powerful flare could
release a huge amount of energy, which could be as high as
$\sim$10$^{32}$~erg in a few tens of minutes via magnetic
reconnection \citep{Priest02,Shibata11,Jiang21,Yan22}. The released
energy rapidly heats the local plasma to higher temperatures such as
dozens of MK, it also efficiently accelerates electrons, protons and
ions to the higher energy of tens of keV to GeV
\citep{Krucker08,Warmuth16,Li22b}. When the magnetic configuration
above the corona is closed such as a strong bipolar coronal magnetic
field, the flare is only a smaller portion of the bigger
destabilization of solar atmospheres, referred to as `confined
flare' \citep{Torok05,Song14,Kliem21}. On the contrary, if the
magnetic confinement of a considerable part of the corona is broken
up, the flare might be associated with a coronal mass ejection
(CME), and thus termed as `eruptive flare'
\citep{Lin05,Yashiro06,Wang07}. Generally speaking, the mass of CMEs
ranges from $\sim$10$^{14}$~g to 10$^{16}$~g
\citep{Chen11,Webb12,Lamy19,Yan21}. We should state here that the
CME has its own magnetic driver rather than a simple explosive
result of the simultaneous flare \citep[see,][]{Benz17}. In some
cases, the CME could drive shock waves, accelerating plasmas that
are escaped from the flare region to energetic particles, and then
enhancing electrons, protons and heavy ions from keV to GeV, which
are collectively referred to as solar energetic particles (SEP)
\citep{Reames99,Desai16}. They could seriously disrupt satellites
and may have important effects on communications
\citep{Emslie04,Temmer21}. In a word, the eruptive processes of
solar flares, CMEs, and SEPs are all related to the magnetic
configuration/driver \citep{Forbes96,Lin00,Toriumi19}, for instance,
their energies are major from the magnetic energy conversion by
reconnection \citep{Priest02,Chen11,Li16,Aschwanden17}.

It has been accepted that the stored energy in complex magnetic
fields is impulsively released through reconnection during the solar
flare \citep{Masuda94,Priest02,Chen20,Tan20}. A fraction of released
energies are used to heat local plasmas, while some others can
accelerate the electrons and ions along magnetic field lines that
may be closed or open \citep{Mann06,Cheng18}. During this process,
the released energies are mainly converted into thermal and
nonthermal energies \citep{Su13,Warmuth16b}, while the radiation in
nearly all the electromagnetic spectra increases rapidly and reaches
an apex in a few tens of minutes \citep{Warmuth16,Benz17,Li21r}. In
particularly, the eruptive flare is commonly associated with a CME,
ejecting a great amount of materials with fastest speeds from the
Sun into the heliosphere \citep{Lin03,Temmer21}. In this process,
the magnetic energies are mostly converted to the kinetic and
gravitational potential energies
\citep{Vourlidas00,Emslie12,Ying18}. The CME with the fastest speed
that exceeds the local Alfv\'{e}n speed can drive shock waves
\citep{Webb12,Lu17}. Such CME-driven shock waves might accelerate
particles to a higher energy and produce the so-called SEPs
\citep{Gloeckler94,Reames99,Palmroos22}, which also represents an
important contribution to the global energy budget of solar
eruptions \citep{Emslie04,Aschwanden14}.

With the aid of coordinated observations from multiple telescopes,
the global energy budget of major flares, CMEs, and SEPs have been
evaluated \citep{Emslie04,Emslie12,Motorina20,Yardley22}. They
estimated their energy contents, and concluded that the energies of
flares and CMEs were roughly comparable \citep{Emslie05,Feng13}.
Moreover, the available magnetic free energies released from the
active regions (ARs) were sufficient to power solar flares and CMEs
\citep{Aschwanden14,Aschwanden17,Thalmann15}. A comprehensive
investigation of the energetics of major flares suggested that the
nonthermal energy of accelerated electrons and ions was able to
supply any flare emission across the electromagnetic spectrum
\citep{Emslie04,Emslie12}, and the nonthermal energy was also larger
than the thermal energy in major solar flares
\citep{Aschwanden15,Aschwanden16,Warmuth20}. Previous observations
also found that the majority of flare radiated energy was released
in the longer wavelengths \citep{Kleint16}, for instance, $\sim$70\%
of the radiation is the WL emission \citep{Kretzschmar11}. The X-ray
ultraviolet (XUV) emission was also found to significantly
contribute to the flare radiated energy, i.e., 19\% at 0-27~nm
\citep{Woods04}. The flare energy in the SXR/EUV emission only
amounts to a few percent of the total radiated energy
\citep{Kretzschmar10,Milligan12,Del18,Dominique18}. They still could
provide us a clue to diagnose those important continuum components
of solar flares. In the lower solar atmosphere at optical and UV/EUV
wavelengths, the Ly$\alpha$ emission was dominated in the measured
radiative losses \citep{Milligan14}, but it was still a minor part
of the total flare radiated energy \citep{Kretzschmar13}. Solar
flares are often accompanied by radio bursts, such as Type II, III
or IV bursts, microwave bursts
\citep[e.g.,][]{Mahender20,Lu22,Su22,Yan23}. However, the flare
energy in radio emissions is much smaller than that radiated in
wavebands of X-rays, UV/EUV, and white light, mainly because that
the radio wave is the longest wavelength in the electromagnetic
spectrum, i.e., microwave, decimeter-wave or even metric-wave
\citep[cf.][]{Pasachoff78}. The energetics of confined flares were
also studied \citep{Zhang19,Zhang21a,Cai21}, and they found the
similar results, for instance, the roughly comparison between the
nonthermal energy and the peak thermal energy, the magnetic free
energy was adequate to support the confined flare.

In September 2017, a large amount of major flares occurred in the
active region (AR) NOAA 12673 \citep{Yang17,Chamberlin18}, and a
major flare on 6 September 2017 was the most intense flare of the
solar cycle 24, whose energy realm could be comparable to the
stellar flare \citep{Kolotkov18}. It was well studied due to the wealth
observational data: (I) The flare quasi-periodic pulsations (QPPs) in
multiple wavelengths, i.e., radio/microwave, Ly$\alpha$,
mid-ultraviolet (MUV) Balmer Continuum, SXR/HXR, and $\gamma$-ray
\citep{Kolotkov18,Karlicky20,Li20,Li21}; (II) The fast evolution of
bald patches \citep{Lee21}; (III) The spectral analysis of nonthermal
electrons, protons and ions during the flare impulsive phase
\citep{Lysenko19,Zhang21b}; (IV) The successive flare eruptions and
their relationship with the complex structure of active region NOAA
12673 \citep{Yan18}. However, the energy partition of this powerful
flare, combination with energetics of the followed CME and SEPs have
not been studied in detail. In this paper, we explored the global
energy budget of the major flare, the accompanied CME, and SEPs.

\section{Observations}
On 6 September 2017, a powerful solar flare occurred in the AR NOAA
12673, it was accompanied by the CME and SEPs. The powerful flare
was simultaneously observed at multiple wavelengths by various
space-based instruments, for instance, the Geostationary Operational
Environmental Satellite (GOES), the Hard X-ray Modulation Telescope
(\textit{Insight}-HXMT) \citep{Zhang20}, Konus-Wind \citep{Aptekar95},
the Large-Yield RAdiometer (LYRA) aboard PRoject for OnBoard
Autonomy 2 (PROBA2) \citep{Dominique13}, the Extreme Ultraviolet
Variability Experiment (EVE) \citep{Woods12}, the Helioseismic and
Magnetic Imager (HMI) \citep{Schou12}, and the Atmospheric Imaging
Assembly (AIA) \citep{Lemen12} on board the Solar Dynamics
Observatory (SDO), and the X-ray Telescope (XRT) \citep{Golub07} of
Hinode, as well as estimated by the Flare Irradiance Spectral
Model-Version 2 (FISM2) \citep{Chamberlin20}. The accompanied CME was
measured by C2 and C3 coronagraphs of Large Angle Spectroscopic
Coronagraph (LASCO) \citep{Brueckner95} for the Solar and
Heliospheric Observatory (SOHO) mission. The SEPs with higher
energetic were recorded by the Energetic and Relativistic Nucleon
and Electron experiment (ERNE) \citep{Torsti95} on board SOHO.
Table~\ref{tab_instr} lists these instruments used in this study.

GOES measures the full-disk solar fluxes in SXR~0.1$-$0.8~nm and
0.05$-$0.4~nm with a time resolution of $\sim$2~s. The emission
measure (EM) and the isothermal temperature of SXR-emitting plasmas
can be determined from their ratio \citep{White05}.
\textit{Insight}-HXMT is designed to mainly search for pulsars,
neutron stars and black holes in HXR and $\gamma$-ray channels
\citep{Zhang20}. It also has the capacity to observe the Sun and
provide the solar high-energy spectrum in HXR and $\gamma$-rays with
a time cadence of about 1~s \citep{Liu20}. Konus-Wind also provides
the solar flux and spectrum in HXR and $\gamma$-rays with varying
time cadences, but the total duration of the flare observation is
only about 250~s \citep{Aptekar95}. PROBA2/LYRA measures the solar
irradiance at four channels with a time cadence of 0.05~s
\citep{Dominique13,Dominique18}. Channels~1 and 2 record the solar
irradiance in far-UV (FUV) centered at hydrogen Ly$\alpha$ 121.6~nm
and MUV from 190~nm to 222~nm, while channels~3 and 4 observe the
Sun in the XUV bandpass. Strictly speaking, the XUV bandpass should
be 0.1$-$5$+$17$-$80~nm and 0.1$-$2$+$6$-$20~nm, due to a drop of
the spectral response in the middle of the bandpass of those two
channels \citep{Dolla12,Dominique13}. We here approximate them simply
by 0.1$-$80~nm and 0.1$-$20~nm for easily describing
\citep{Dominique18,Li21}. The EUV SpectroPhotometer (ESP) of SDO/EVE
records the solar irradiance with a time cadence of 0.25~s in one
SXR (0.1-7~nm) band and four EUV wavelengths centered around 19~nm
(17.2$-$20.8~nm), 25~nm (23.1$-$27.8~nm), 30~nm (28.0$-$31.8~nm),
and 36~nm (34.0$-$38.7~nm) \citep{Didkovsky12}. The Multiple EUV
Grating Spectrographs (MEGS) of SDO/EVE measures the solar spectral
irradiance in EUV wavelengths from 6 to 106 nm plus the
Ly$\alpha$~121.6~nm \citep{Hock12,Woods12}. The temporal evolutions
of EUV and Ly$\alpha$ emissions are provided by merging the spectral
measurement, which have a time cadence of 10~s.

SDO/HMI can provide the continuum images, the line-of-sight (LOS)
and vector magnetograms with time cadences of 45/720~s
\citep{Schou12}. SDO/AIA captures solar images in multiple
wavelengths nearly simultaneously, containing seven EUV and two UV
bands \citep{Lemen12}. Their time cadences are 12~s for EUV
observations and 24~s for UV images. The HMI and AIA images have a
spatial pixel size of 0.6$^{\prime\prime}$ after a standard
correction, that is, aia\_prep.pro or hmi\_prep.pro. Hinode/XRT
provides solar coronal images at X-ray channels \citep{Golub07}. In
this study, a XRT image at the Be\_med channel is used, which has a
spatial scale of about 1.03$^{\prime\prime}$ per pixel. FISM2 is an
empirical model of the solar spectral irradiance in X-ray and EUV/UV
wavelengths, which is created to fill the actual observational gap.
It covers the wavelength range from 0.01~nm to 190~nm with a spectral
bin of 0.1 nm, and a time cadence of 60~s \citep{Chamberlin20}.

The C2 and C3 coronagraphs on aboard SOHO/LASCO \citep{Brueckner95}
are used to measure the solar corona in WL continuum images from
1.1R$_{\odot}$ to 32R$_{\odot}$. The higher energetic particles
observed by the SOHO/ERNE \citep{Torsti95} are applied to measure the
SEP energy, and we can only trust the proton spectrum above 20~MeV.

\section{Energy partition in the powerful flare}
\subsection{Overview of the solar flare}
\begin{figure}[ht]
\centering
\includegraphics[width=0.6\linewidth,clip=]{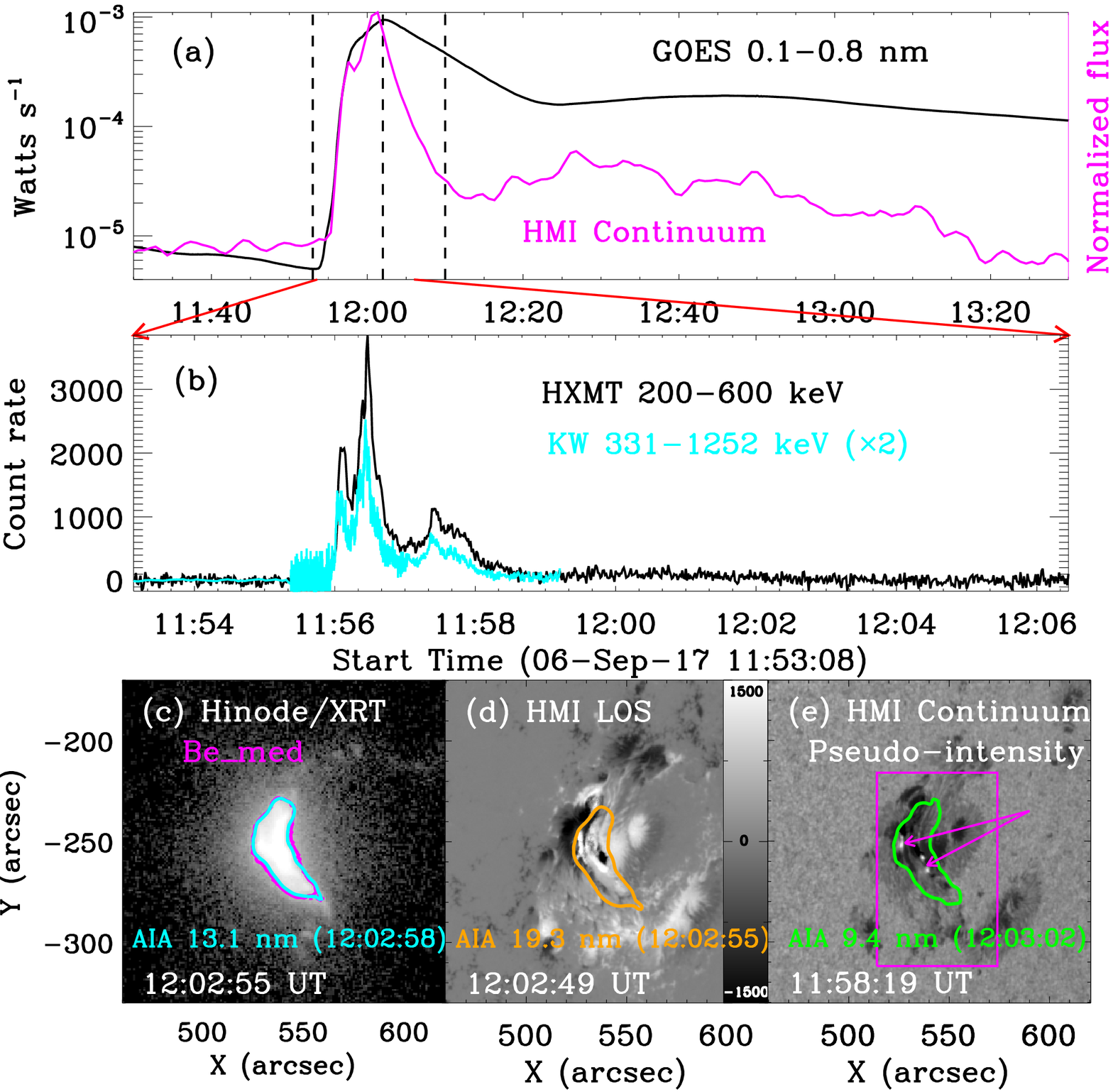}
\caption{Overview of the powerful flare on 6 September 2017.
Panels~(a) and (b): Multi-wavelength light curves measured by GOES,
SDO/HMI, HXMT, and Konus-Wind. Panel~(c): Snapshot observed by
Hinode/XRT, the overplotted magenta contour represents the SXR
emission at a level of 1/e$^2$. Panel~(d): LOS magnetogram measured
by SDO/HMI. Panel~(e): Pseudo-intensity image obtained by SDO/HMI
continuum data, two magenta arrows point out the locations where the
while light enhancements took place. The overplotted cyan, orange,
and green contours are made from SDO/AIA emissions. The magenta
rectangle marks the integrated region for the light curve in white
continuum. \label{img}}
\end{figure}

The major flare on 6 September 2017 was the most powerful flare
during the solar cycle 24. It was an X9.3 class according to the
GOES SXR flux in 0.1$-$0.8~nm, as shown in Figure~\ref{img}~(a). The
X9.3 flare began to enhance at about 11:53~UT, reached its maximum
at $\sim$12:02~UT, and stopped at about
12:10~UT\footnote{https://www.solarmonitor.org/?date=20170906}, as
indicated by the dashed vertical lines. Panel~(b) plots the HXR and
$\gamma$-ray light curves during the impulsive phase of the X9.3
flare, which were measured by the \textit{Insight}-HXMT at
200$-$600~keV and Konus-Wind at 331$-$1252~keV, respectively. Noting
that the Konus-Wind light curve has been multiplied by 2.0, so it
can be clearly seen and compared with the \textit{Insight}-HXMT flux
in a same panel. They both reveal a series of regular and periodic
pulsations, suggesting the existence of flare-related QPPs in HXR
and $\gamma$-ray emissions \citep{Li20,Li21}, and also implying the
presence of higher energetic electrons/ions in the X9.3 flare
\citep{Lysenko19,Motorina20,Zhang21b}. However, the observational
time of Konus-Wind is much shorter than that of
\textit{Insight}-HXMT, so only the \textit{Insight}-HXMT spectrum is
analyzed in this study.

Figure~\ref{img}~(c)$-$(e) presents the multi-wavelength images with
a same field-of-view (FOV) of
$\sim$160$^{\prime\prime}$$\times$160$^{\prime\prime}$ at around the
flare peak time, the color contours are made from the AIA
high-temperature EUV images, such as AIA~13.1~nm (cyan), 19.3~nm
(orange) and 9.4~nm (green). Panel~(c) shows the SXR image measured
by the Hinode/XRT at Be\_med channel, and the overlaid magenta
contour represents the SXR emission source that is greater than the
$\frac{1}{e^2}$ of maximal source intensity \citep{Zhang21b}, which
is consistent well with the SDO/AIA emission in high-temperature EUV
wavelengths, for instance, the hot flare loops marked by the color
contours match well with each other. It is used to estimate the
flare area ($A$) of hot (or SXR-emitting) plasmas. In this study, we
refer to the width/diameter of laser beams, which could be defined
at points where the intensity decreases to $\frac{1}{e^2}$ of the
maximum intensity. Panel~(d) presents the LOS magnetogram observed
by HMI, showing strong and complex magnetic fields underlying the
hot flare loop. Panel~(e) draws a pseudo-intensity image derived
from HMI continuum filtergrams, and the WL radiation is enhanced, as
shown by the magenta arrows. Thus, the weak white light flare (WLF)
can be easily detected from the pseudo-intensity image
\citep{Song18,Li23}. The light curve integrated over the flare
region (magenta rectangle) is also given in panel~(a), which clearly
show an enhancement during the flare impulsive phase, confirming
that it is a WLF. On the other hand, when the HMI pseudo continuum
is used as a proxy for the WL continuum, it works well in the quiet
sun, but could be significant deviations in solar flares
\citep{Svanda18}. Therefore, we do not use it for further
calculation.

\begin{figure}[ht]
\centering
\includegraphics[width=0.6\linewidth,clip=]{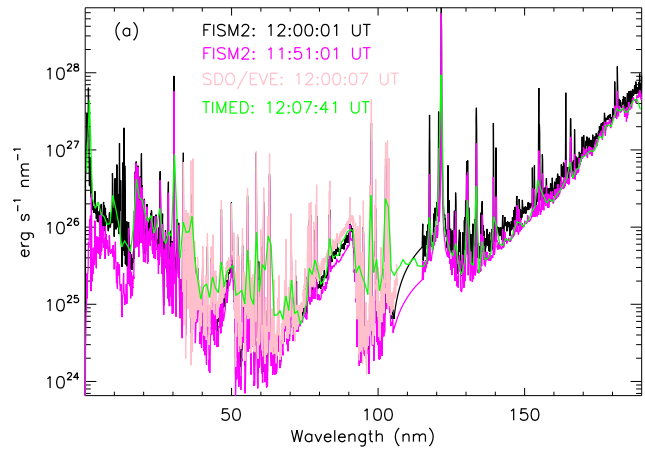}
\caption{Solar spectra between 0.01$-$190~nm derived from FISM2
before (magenta) and during (black) the powerful flare. The
overplotted spectra are measured by SDO/EVE (pink) and TIMED
(green), respectively. \label{spec}}
\end{figure}

Figure~\ref{spec} shows the solar spectrum from 0.01~nm to 190~nm
derived from FISM2 with a spectral bin of 0.1~nm. The black and
magenta line profiles represent the flare and quiet-Sun spectra, for
instance, during the X9.3 flare and before the flare onset time. The
flare spectrum from 33.35$-$106.59~nm measured by the SDO/EVE is
also overplotted by the pink line, it has a time cadence of 10~s and
a spectral bin of 0.02~nm \citep{Hock12,Woods12}. The flare spectrum
(green) between 0.5$-$190~nm with a lower spectral resolution of
1~nm is observed by the Thermosphere Ionosphere Mesosphere
Energetics Dynamics (TIMED) at about 12:07:41~UT, but it only lasts
for about 3~minutes \citep{Woods05}. Here, the solar spectrum
observed by TIMED is used to correct the FISM2 model data, which
demonstrates to match well with the observational data, as shown in
Figure~\ref{spec}. For easier description, we simply divided the
observed wavelength ranges into HXR ($<$0.1~nm), SXR (0.1$-$7~nm),
EUV (7$-$120~nm), FUV (120$-$190~nm) and MUV (190$-$222~nm)
\citep{Del18}, while the XUV contains SXR and EUV wavelength ranges
at 0.1$-$20~nm and 0.1$-$80~nm recorded by LYRA \citep{Dominique18}.

\begin{table*}
\centering
\caption{Instruments used in this study.} \label{tab_instr}
\tabcolsep 10pt
\begin{tabular*}{\textwidth}{ccccc}
\hline\hline
Instruments & Channels       &  Time cadence (s) & Wavebands  & Description   \\
\hline
           &  0.05$-$0.4~nm   &  $\sim$2.0     &  SXR     &      Flux \\
GOES       &  0.1$-$0.8~nm    &  $\sim$2.0     &  SXR     &      Flux \\
\hline
\textit{Insight}-HXMT  &  200$-$600~keV   &   1.0       &  HXR/$\gamma$ & Spectrum \\
\hline
Konus-Wind &  331$-$1252~keV  & 0.256$-$8.192  &  $\gamma$-ray & Spectrum \\
\hline
           &  0.1$-$20~nm      &  0.05       &  XUV   &  Flux (gap: 2$-$6~nm)  \\
PROBA2/LYRA  &  0.1$-$80~nm      &  0.05       &  XUV   & Flux  (gap: 5$-$17~nm) \\
           &  120$-$123~nm     &  0.05       &  Ly$\alpha$ & Flux  \\
           &  190$-$222~nm     &  0.05       &  MUV      & Flux  \\
\hline
          &  0.1$-$7~nm         & 0.25       &  SXR    &   Flux    \\
          &  17.2$-$20.8~nm     & 0.25       &  EUV    &   Flux    \\
          &  23.1$-$27.8~nm     & 0.25       &  EUV    &   Flux  \\
SDO/EVE/ESP &  28.0$-$31.8~nm     & 0.25       &  EUV    & Flux    \\
SDO/EVE/MEGS-B & 33.3$-$107~nm    &  10        &  EUV    &   Spectrum \\
SDO/EVE/MEGS-P & 121.6~nm         &  10        &  Ly$\alpha$ & Flux \\
           &   617.3~nm           & 45/720     &  LOS/Vector & Magnetograms  \\
SDO/HMI    &   617.3~nm           & 45         &  Continuum  & image \\
SDO/AIA    &  9.4, 13.1, 19.3~nm  & 12         &  EUV    & image \\
\hline
Hinode/XRT &  Be\_med            & --          &  SXR     &  image   \\
\hline
FISM2        & 0.01-190~nm       &  60         &  Empirical &  Spectrum \\
\hline
SOHO/LASCO/C2  &    --     &     --        &  WL &  CMEs \\
SOHO/LASCO/C3  &    --     &     --        &  WL &  CMEs \\
\hline
SOHO/ERNE      &    --     &     --         &  $>$20~MeV & SEPs \\
\hline
\end{tabular*}
\end{table*}

\subsection{Flare energy radiated in multiple wavebands}
In this section, we focus on energy contents of the X9.3 flare
radiated in multiple wavelengths shortward of $\sim$222~nm, for
instance, the radiated energies in SXR, XUV, EUV, Ly$\alpha$, and
MUV. To obtain them, we firstly remove the background emission from
the solar observational irradiance. The background emission is
defined as the solar irradiance before the flare onset time ($t_o$).
Briefly, some data points before $t_o$ are extracted and then
performed a linear fitting \citep{Zhang19}. Then, we can calculate
the radiated energy ($U_\lambda$) in the specific waveband
($\lambda$) by integrating the background-subtracted flux
($F_\lambda$) over the flare time interval
\citep{Emslie12,Milligan14,Zhang19,Zhang21a,Cai21}, as shown in
equation~\ref{eq_wav}.
\begin{equation}
U_\lambda=2 \pi d^2 \int_{t_1}^{t_2} F_\lambda(t)\, dt,
\label{eq_wav}
\end{equation}
\noindent Here, $d$ represents the distance between the Sun and the
Earth, which is 1~AU. $t_1$ and $t_2 $ are the began and stop time
of the X9.3 flare in a certain waveband. It should be pointed out
that $t_1$ and $t_2 $ are a bit different in various channels,
mainly due to the observational fact that certain-waveband radiation
could be dominated in different flare phases.

\begin{figure}[ht]
\centering
\includegraphics[width=0.6\linewidth,clip=]{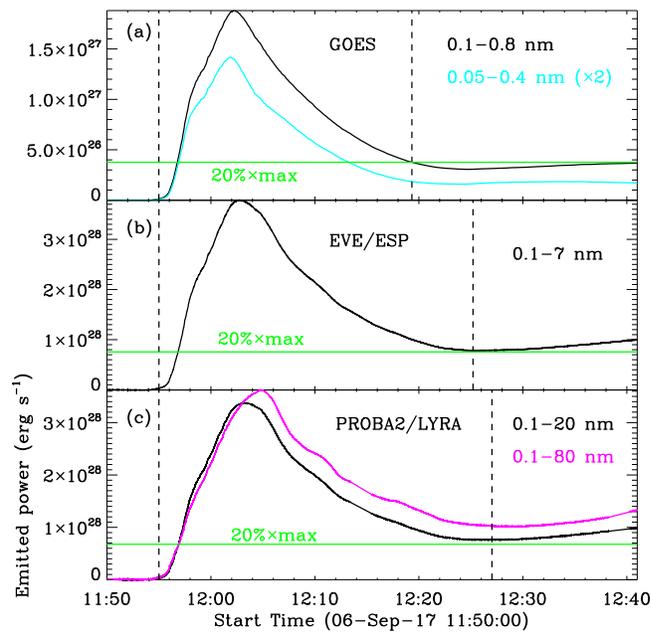}
\caption{Background-subtracted SXR/XUV fluxes in 0.05$-$0.4~nm,
0.1$-$0.8~nm, 0.1$-$7~nm, 0.1$-$20~nm, and 0.1$-$80~nm. The vertical
dashed lines marks their integrated time intervals ($t_1$ and $t_2
$), and the horizontal green lines represent the 20\% of their
maximum (the black curve in each panel). \label{en_sxr}}
\end{figure}

First of all, we calculate the flare radiated energy in SXR
(0.05$-$0.4~nm, 0.1$-$0.8~nm, and 0.1$-$7~nm) and XUV (0.1$-$20~nm,
0.1$-$80~nm) wavelength ranges measured by GOES, SDO/EVE, and
PROBA2/LYRA, respectively. Figure~\ref{en_sxr} presents the light
curves of those five channels after removing their background
emissions. It can be seen that their temporal profiles are quite
similar, but their peaks appear to slightly delay in time. This is
consistent with the observational fact that cooler wavelengths
usually peak after the hottest ones in solar flares. That is, the
flare radiation in the longer wavelength that has lower temperatures
will spend much time to cool down \citep{Kretzschmar13}. The begin
time ($t_1$) for integrating is defined as the time when the flare
starts to increase, as marked by the dashed vertical lines. It is at
about 11:55:00~UT, and the GOES SXR fluxes have a 1-s deviation
because the time cadence of GOES is $\sim$2~s. The stop time ($t_2$)
for integrating is a little complicated. In Figure~\ref{img}~(a),
the GOES SXR flux in 0.1$-$0.8~nm reveals a second peak at roughly
12:40~UT, which is after the major flare and could affect its stop
time. Therefore, $t_2$ is defined as the time when the flux drops to
20\% of its peak value, as indicated by the green lines in
panels~(a) and (b). However, panel~(c) shows that the XUV flux does
not drop to 20\% before increasing of the second peak. So, $t_2$ is
the XUV flux decreases to its valley value, as indicated by the
dashed vertical line in panel~(c). Then using equation~\ref{eq_wav},
the flare radiated energy in SXR and XUV wavebands is estimated, as
listed in Table~\ref{tab_rad}. The flare radiated energy in GOES
SXR~0.1$-$0.8~nm is about 1.4$\times$10$^{30}$~erg, which is 3~times
larger than the short-wavelength GOES channel, but it is much
smaller than that radiated in ESP~0.1$-$7~nm, such as
3.4$\times$10$^{31}$~erg. On the other hand, the flare radiated
energies in two XUV channels measured by LYRA are roughly equal,
which are in the order of magnitude 10$^{31}$~erg. We want to stress
that the definition of ($t_2$) could slightly affect the estimation
of the radiated energy, i.e., about 2\% \citep{Zhang19}, but it is
hardly influence for the order of magnitude.

\begin{table}
\centering
\caption{Radiated energy of the X9.3 flare in the specifical waveband.}
\label{tab_rad}
\tabcolsep 8pt 
\begin{tabular}{cccc}
\hline\hline
 Wavebands        &  t$_1$ (UT)& t$_2$ (UT)&    Energy (erg)      \\
\hline
U$_{0.05-0.4}$    &  11:54:59 & 12:19:19 &  4.8$\times$10$^{29}$ \\
U$_{0.1-0.8}$     &  11:54:59 & 12:19:19 &  1.4$\times$10$^{30}$ \\
U$_{0.1-7}$       &  11:55:00 & 12:25:12 &  3.4$\times$10$^{31}$ \\
U$_{0.1-20}$      &  11:55:00 & 12:27:01 &  3.1$\times$10$^{31}$ \\
U$_{0.1-80}$      &  11:55:00 & 12:27:01 &  3.5$\times$10$^{31}$ \\
U$_{17.2-20.8}$   &  11:53:57 & 12:27:56 &  4.8$\times$10$^{29}$ \\
U$_{23.1-27.6}$   &  11:53:57 & 12:27:56 &  4.0$\times$10$^{29}$ \\
U$_{28.0-31.8}$   &  11:53:57 & 12:27:56 &  8.2$\times$10$^{29}$ \\
U$_{33.3-61.0}$   &  11:53:57 & 12:14:07 &  3.1$\times$10$^{29}$ \\
U$_{61.0-79.1}$   &  11:53:57 & 12:14:07 &  1.8$\times$10$^{29}$ \\
U$_{79.1-107}$    &  11:53:57 & 12:14:07 &  8.9$\times$10$^{29}$ \\
U$_{190-222}$     &  11:55:16 & 12:07:50 &  1.5$\times$10$^{30}$ \\
U$_{121.6}$       &  11:55:17 & 12:07:47 &  1.1$\times$10$^{30}$ \\
\hline\hline
\end{tabular}
\end{table}

\begin{figure}[ht]
\centering
\includegraphics[width=0.6\linewidth,clip=]{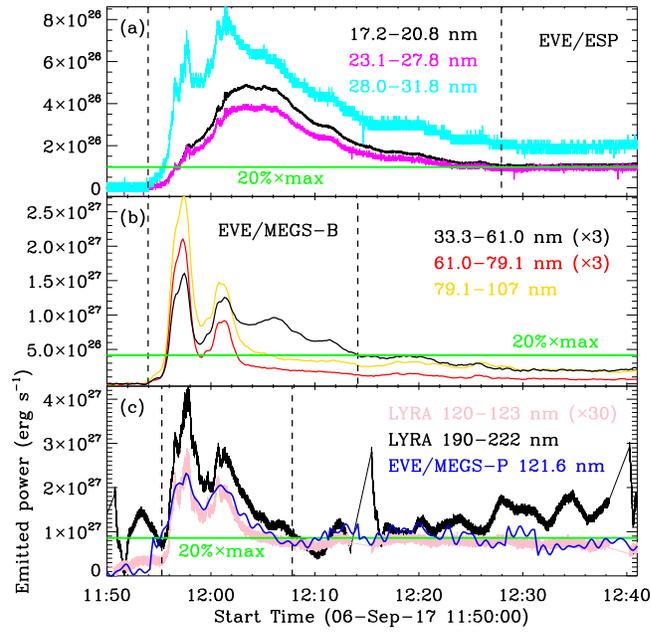}
\caption{Background-subtracted EUV/MUV light curves in
17.2$-$20.8~nm, 23.1$-$27.8~nm, 28.0$-$31.8~nm, 33.3$-$61.0~nm,
61.0$-$79.1~nm, 79.1$-$107~nm, 120$-$123~nm, 190$-$222~nm, and
121.6~nm. The vertical dashed lines outline their integrated time
intervals ($t_1$ and $t_2 $), and the horizontal green lines
represent the 20\% of their maximum (the black curve in each panel).
\label{en_euv}}
\end{figure}

Like in SXR and XUV channels, we calculate the EUV/MUV radiated
energy of the X9.3 flare recorded by SDO/EVE and LYRA, as shown in
Figure~\ref{en_euv}. We plot the light curves after removing their
background emissions in EUV, Ly$\alpha$, and MUV wavebands, which
are recorded by ESP and MEGS on board SDO/EVE, and LYRA,
respectively. Here, some light curves have been multiplied by a
fixed factor to show them clearly in the same window. $t_1$ is
defined as the fixed time instance of the flare starting to increase
in the chosen waveband, and $t_2$ is determined from the time when
the radiated flux decrease to 20\% of its peak value, as indicated
by the dashed vertical lines. Then the radiated energy in EUV
wavelengths can be estimated with equation~\ref{eq_wav}. The details
can be seen in Table~\ref{tab_rad}. We can find that the flare
radiated energy in the EUV wavelength is roughly in the order of
10$^{29}$~erg, which is one order less than that in the MUV and
Ly$\alpha$ wavebands ($\sim$10$^{30}$~erg). They are much less (two
order) than the radiated energy in SXR/XUV channels (10$^{31}$~erg).
This is consistent with our previous result, for instance, the
radiated energy in SXR~0.1$-$7~nm recorded by ESP is similar to that
in XUV~0.1$-$20~nm and 0.1$-$80~nm measured by LYRA, as listed in
the fourth column of Table~\ref{tab_rad}. We wanted to state that
the flare energy released in the Ly$\alpha$ channel is estimated
from the SDO/EVE observation, similar to the previous result
\citep[cf.][]{Milligan14}. On the other hand, The factor $\sim$30
between SDO/EVE and LYRA is of the same order than the one found by
\cite{Wauters22} with respect to the Ly$\alpha$ channel of GOES and
LYRA.

\subsection{Radiative loss energy and peak thermal energy}
\begin{figure}[ht]
\centering
\includegraphics[width=0.6\linewidth,clip=]{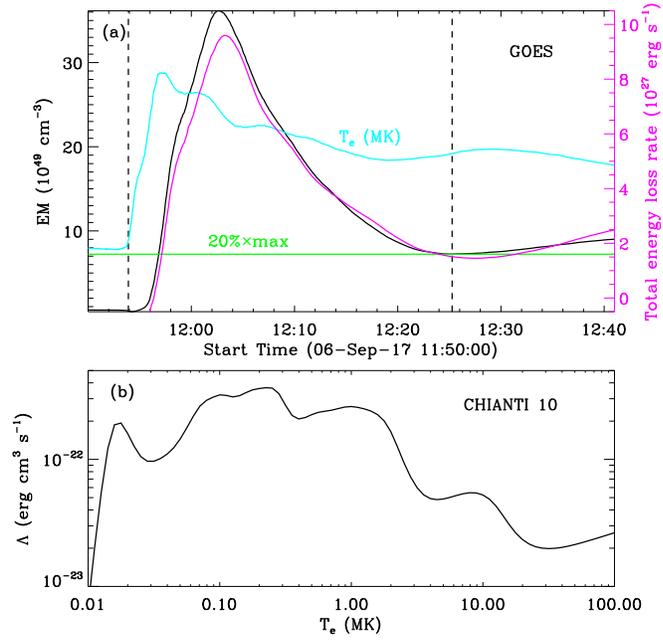}
\caption{Panel~(a): Temporal evolutions of EM (black), temperature
(cyan), and total energy loss rate (magenta) obtained from the GOES
SXR observation. The vertical dashed lines outline the integrated
time interval, and the horizontal green lines represent the 20\% of
the peak EM. Panel~(b): Radiative loss rate as a function of the
temperature derived from CHIANTI 10. \label{en_loss}}
\end{figure}

In this section, we calculate the radiative loss energy ($U_{rad}$)
and peak thermal energy ($U_{pth}$) of SXR-emitting plasmas based on
the GOES and Hinode/XRT observational data. Assuming the optical
thin radiation, $U_{rad}$ can be calculated with
equation~\ref{eq_rad}.
\begin{equation}
U_{rad} = \int_{t_1}^{t_2} \Lambda (T_e (t)) \times {\rm EM}(t)\, dt,
\label{eq_rad}
\end{equation}
\noindent Where $\Lambda (T_e)$ is the radiative loss rate as a
function of $(T_e)$, which can be obtained from the CHIANTI 10
database \citep{Delz21}, as shown in Figure~\ref{en_loss}~(b). EM
and $(T_e)$ denote the emission measure and electron temperature,
which are derived from two GOES SXR fluxes using an isothermal
assumption \citep{White05}, as shown in Figure~\ref{en_loss}~(a).
$t_1$ and $t_2$ represent the integrated time obtained from the
temporal profile of EM evolution, as indicated by the two dashed
vertical lines. Then, the radiative loss energy from SXR-emitting
plasmas is estimated to 8.7$\times$10$^{30}$~erg.
Figure~\ref{en_loss}~(a) also draws the temporal evolution of the
total energy loss rate (magenta) after removing the background,
which is provided by the GOES team. We can calculate the total
energy loss ($U_{trad}$) of hot plasmas by integrating over the
background-subtracted total energy loss rate. It is estimated to
about 9.0$\times$10$^{30}$~erg, which is roughly equal to $U_{rad}$,
as can be seen in Table~\ref{tab_flare}. The radiative loss energy
of the X9.3 flare is 6 times larger than the radiation in
GOES~0.1$-$0.8~nm, which agrees with previous observational results,
for instance, one order of magnitude deviations
\citep{Emslie12,Feng13,Zhang19}.

Using equation~\ref{eq_pth}, the peak thermal energy ($U_{pth}$) from
SXR-emitting plasmas can be estimated.
\begin{equation}
U_{pth} \approx 3 k_b T_e \sqrt{{\rm EM} \times V}, \label{eq_pth}
\end{equation}
\noindent Here, $k_b$ denotes the Boltzmann constant. $V$ represents
the volume of SXR-emitting plasmas, which can be estimated from the
SXR image at the flare peak time observed by Hinode/XRT. The magenta
contour in Figure~\ref{img}~(c) outlines the major flare region contains
the SXR emission at a level of $\frac{1}{e^2}$, which could be
identified as the flare area ($A$) of SXR-emitting plasmas after
correcting the projection effect \citep{Zhang19}, such as
$\sim$690~Mm$^2$. The volume can be expressed as
$V=A^{\frac{3}{2}}$. Obviously, the flare area is important for
estimating the peak thermal energy, since it is directly related to
the volume of SXR-emitting plasmas. In this study, we also present
the SDO/AIA emissions in high-temperature EUV wavelengths (13.1~nm,
19.3~nm, 9.4~nm), as indicated by the overplotted cyan, orange, and
green contours in Figure~\ref{img}~(c)$-$(e). Those SDO/AIA emissions
include hot plasmas of the solar flare, and they match well with the
SXR emission observed by Hinode/XRT, confirming the flare area of
SXR-emitting plasmas. Then, $U_{pth}$ reaches its maximum when $T_e
\sqrt{EM}$ is maximal, which is $\sim$3.1$\times$10$^{31}$~erg.
Noting that we assume a volumetric filling factor of 1.0
\citep[e.g.,][]{Emslie12,Zhang19}.

\subsection{Nonthermal energy in flare-accelerated ions}
\begin{figure}[ht]
\centering
\includegraphics[width=0.6\linewidth,clip=]{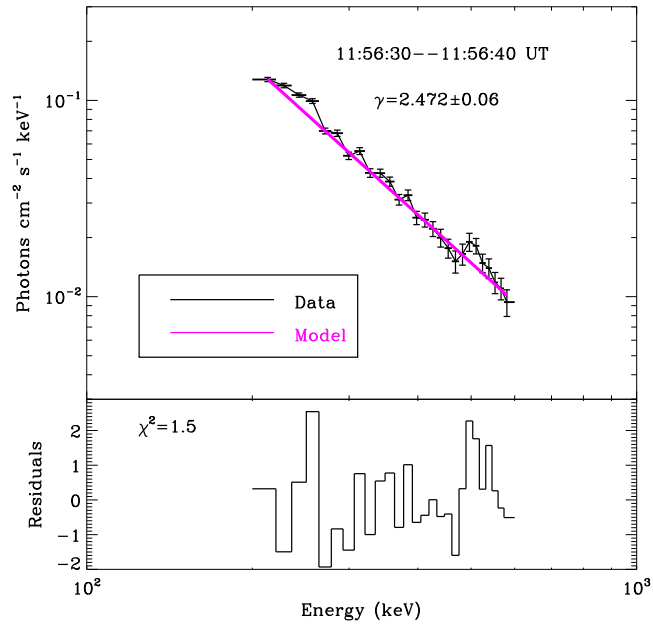}
\caption{The \textit{Insight-HXMT} spectrum with an integrated time
of 10~s during $\sim$11:56:30$-$11:56:40~UT, the magenta line
represents the fitting model. The spectral index ($\gamma$) and the
Chi-squared residual ($\chi^2$) are labeled. \label{hxr}}
\end{figure}

The HXR spectrum produced by solar flare often reveals the power-law
distribution with low-energy and high-energy cutoffs, which could be
a useful diagnostic of energetic electrons, protons and ions
\citep{White11}. However, the HXR spectrum below 100~keV is short of
observations during the impulsive phase of the X9.3 flare
\citep{Lysenko19}. Therefore, we focus on the HXR spectrum in energy
ranges between 200$-$600~keV, which is observed by
\textit{Insight}-HXMT\footnote{http://hxmtweb.ihep.ac.cn/documents/497.jhtml},
as shown in Figure~\ref{img}~(b). Then, we perform a power-law model
to fit the HXR spectrum \citep{Zhang21b}. Figure~\ref{hxr} shows the
HXR spectrum (black) and its fitting result (magenta) at the main
peak of the \textit{Insight}-HXMT light curve during
11:56:30$-$11:56:40~UT. The observational HXR spectrum and the
fitting model appear to match well. Moreover, The Chi-squared
residual ($\chi^2=1.5$) also implies a reasonable fitting result,
i.e., $\chi^{2}<3$ \citep{Sadykov15}. Thus, we can obtain the
spectral photon index, which is about 2.472$\pm$0.063.

Next, we could estimate the nonthermal power ($P$) above a cutoff
energy (E$_c$) for energetic electrons and protons with equation
\ref{eq_non} \citep[cf.][]{Zhang16,Li18,Li22}.
\begin{equation}
 P (E \geq E_c) = 1.16 \times 10^{24} \gamma^{3} I_1 (\frac{E_c}{E_1})^{-(\gamma -1)},
 \label{eq_non}
\end{equation}
\noindent Where E$_1$ is the lower cutoff energy. I$_1$ denotes the
photon count rates, which has a range of
10$^1-$10$^5$~photon~s$^{-1}$~cm$^{-2}$ at energies of
E~$\geq$~20~keV and spectral indexes of $\sim$3. For the X9.3 flare,
we can estimate the nonthermal energy above 200~keV by integrating
the nonthermal power over a specified time interval between
11:56:30~UT and 11:56:40~UT. Using the same method, we obtain the
total nonthermal energy of energetic electrons/protons from 11:56~UT
to 11:59~UT during the X9.3 flare, which is about
(1.5$\pm$0.2)$\times$10$^{31}$~erg. At last, we estimate the
nonthermal energy (E$_{nth}$) of all accelerated ions above 200~keV
by assuming that the total ion energy could be three times larger
than the electron/proton energy \citep{Emslie12,Aschwanden17}, for
instance, E$_{nth} \sim$4.5$\times$10$^{31}$~erg, as listed in
Table~\ref{tab_flare}. As \textit{Insight}-HXMT does not measure the
entire duration of the HXR/$\gamma$-ray emission, our measurement of
the nonthermal energy is indeed a minimum estimation.

\begin{table}
\centering \caption{Energy components for the X9.3 flare.} \label{tab_flare}
\tabcolsep 5pt 
\begin{tabular}{ccccc}
\hline\hline
U$_{rad}$ & U$_{trad}$ &  U$_{pth}$ & E$_{nth}$  \\
\hline
8.7$\times$10$^{30}$~erg &  9.0$\times$10$^{30}$~erg & 3.1$\times$10$^{31}$~erg  & $\geq$4.5$\times$10$^{31}$~erg  \\
\hline\hline
\end{tabular}
\end{table}

\section{Energetics of the accompanied CME}
Here, we first calculate basic quantities such as mass, height, and
speed, and then derive the gravitational potential and kinetic
energies of the accompanied CME. Traditionally, the mass is
estimated according to the Thomson scattering theory by assuming
that all of the emission along a given LOS comes from electrons
located on the Thomson Sphere (where the scattering angle equals to
90$^{\circ}$ and the Thomson scattering is the most efficient). For
the FOV ($<$ 30$R_{\odot}$) of the LASCO/C2 and C3, the Thomson
Sphere virtually coincides with the plane of sky (POS), which means
that the traditional method is only reliable for CMEs that propagate
along or close to the POS. In the case of CMEs propagating away from
the POS, especially Halo CMEs, the Thomson scattering drastically
drops and more electrons are expected to reproduce the observed
white-light intensity, which increases the uncertainties in the mass
determinations. However, such uncertainties can be partly reduced by
assuming that all the electrons contributing to the white-light
emission lies on a different angle than the POS \citep{Vourlidas10}.
This angle should be the propagation angle of the CME with respect
to the POS.

\begin{figure}[ht]
\centering
\includegraphics[width=0.6\linewidth,clip=]{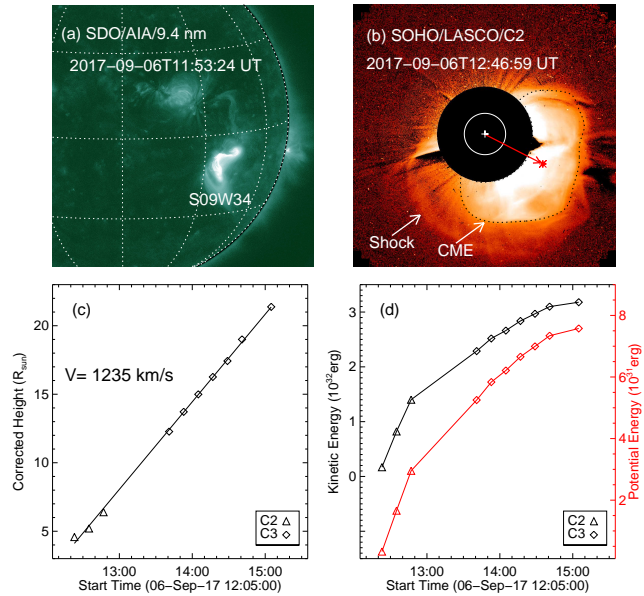}
\caption{Physical properties of the CME on 6 September 2017.
Panel~(a): Solar disk image taken by SDO/AIA at 9.4 nm. (b) Coronal
image taken by SOHO/LASCO/C2. The small white circle inside the
occulter represents the solar disk, with the``+" symbol indicating
the solar center. The dotted line outlines the CME projected on the
plane of sky. The ``$\ast$" symbol indicates the projected location
of the CME mass center. (c) Height-time profile of the CME mass
center. A linear fit of the profile reveals an average speed. (d)
Evolution of kinetic (black) and potential (red) energies of the
CME. The ``$\triangle$'' and ``$\diamond$'' symbols show
measurements from LASCO/C2 and C3, respectively. \label{en_cme}}
\end{figure}

In our case, as seen in Figure~\ref{en_cme}~(a) and (b), the CME
under study originates from active region NOAA 12673 located at
about 56 degree from the west solar limb. Assuming that the CME
propagates radially from the source active region
\citep{Reiner03,Kahler05}, the CME angle of propagation with respect
to the POS is estimated to be about 56 degree. Base on this angle
and the Thomson scattering formulation \citep{Billings66}, the
coronal images containing the CME, after subtracting a suitable
pre-event image, can be converted to the corresponding mass images.
The total mass of the CME ($M_{cme}$) is computed by summing the
masses in the pixels encompassed in the CME, as delineated by the
black dotted line in Figure~\ref{en_cme}~(b). The radial height
($R_{mc}$) of the mass center of the CME from the solar center is
given by equation~\ref{eq_r}.
\begin{equation}
R_{mc}=\frac{\sum m_{i}r_{i}}{\sum m_{i}},
\label{eq_r}
\end{equation}
\noindent Here $m_i$ is the mass in each pixel, r$_i$ is the
corrected POS height of each pixel (POS height divided by cosine of
the CME propagation angle). We calculate $R_{mc}$ for each of the
coronal images as the CME propagated through the field of view of
LASCO. Figure~\ref{en_cme}~(c) shows the evolution of the height of
the CME mass center. A linear fit to the height-time data reveals an
averaged velocity ($v_{cme}$) of $\sim$1235~$\rm{km~s}^{-1}$ for the
CME. The calculation of the height and speed as described above has
an advantage of involving only the measurement of the CME mass
center. Based on the computed mass, height, and velocity, the
potential ($E_p$) and kinetic ($E_k$) energies of the CME can be
simply estimated by equation~\ref{eq_cme}.
\begin{eqnarray}
E_p&=&GM_{\odot}M_{cme}(\frac{1}{R_{\odot}}-\frac{1}{R_{mc}}),  \\
E_k&=&\frac{1}{2}M_{cme}v_{cme}^2,
\label{eq_cme}
\end{eqnarray}
\noindent Where $G$ is the gravitational constant, $M_{\odot}$ and
$R_{\odot}$ represent the mass and radius of the Sun. Here, we
measure the potential energy relative to the solar surface, and the
obtained results are illustrated in Figure~\ref{en_cme}~(d). As can
be seen, both the kinetic energy and the potential energy of the CME
increase with time (and hence the altitude). The kinetic energy is
obvious larger than the gravitational potential energy, and the
total CME energy is about 4$\times$10$^{32}$~erg. Note that, the CME
is partially occulted by the coronagraph mask at early times, which
leads to an underestimation of the CME energy.

\section{Magnetic free energy}
\begin{figure}[ht]
\centering
\includegraphics[width=0.6\linewidth,clip=]{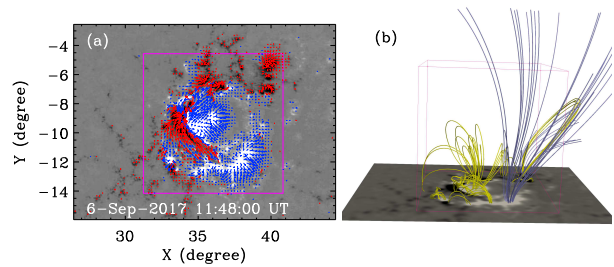}
\caption{Panel~(a): SDO/HMI vector magnetogram of the AR NOAA 12673
hosting the flare and CME. Panel~(b): Nonpotential magnetic
configurations derived from the NLFFF extrapolation. The yellow and
purple lines represent the closed and open magnetic field lines,
respectively. The magenta box outlines the region that is used to
calculate the magnetic free energy.\label{mag_nlfff}}
\end{figure}

In order to estimate the magnetic free energy stored in the AR NOAA
12673, we perform a nonlinear force-free field (NLFFF) extrapolation
using the `weighted optimization' method
\citep{Wheatland00,Wiegelmann12} after preprocessing the photospheric
boundary to meet the force-free condition \citep{Wiegelmann06}, as
shown in Figure~\ref{mag_nlfff}. This method uses the photospheric
magnetic field vector (panel~a) observed by the SDO/HMI and the
potential field derived from the vertical component of the magnetic
field with a Green's function algorithm as a boundary condition to
reconstruct three-dimensional magnetic fields. For the
extrapolation, we bin the data to 1.0$^{\prime\prime}$~pixel$^{-1}$
and adopt a computation box of 300$\times$224$\times$224 uniform
grid points (225$\times$168$\times$168~Mm$^3$). Panel~(b) presents
the NLFFF extrapolated results, and it shows both the closed and
open magnetic field lines, but the open magnetic field lines prefer
to appear at the non-flare area. The magnetic free energy
(E$_{free}$) can be determined from the NLFFF (E$_{nl}$) energy and
the potential field energy (E$_{pf}$) with equation~\ref{eq_mag}

\begin{eqnarray}
  E_{free} = E_{nl} - E_{pf} = \int_{V} \frac{B_{nl}^2 - B_{pf}^2}{8\pi} \, dV,
   \label{eq_mag}
\end{eqnarray}
\noindent Here, B$_{nl}$ and B$_{pf}$ represent the magnetic
strength of the nonpotential and potential fields, which can be
obtained from the NLFFF extrapolation. $V$ stands for the coronal
volume of the AR, and a value of about
120$\times$120$\times$120~Mm$^3$ is taken into consideration here,
as indicated by the magenta box in Figure~\ref{mag_nlfff}. Thus, the
magnetic free energy can be estimated to about
2$\times$10$^{33}$~erg.

\section{Energy of SEPs}
\label{sec:seps} 
\begin{figure}[ht]
\centering
\includegraphics[width=0.6\linewidth]{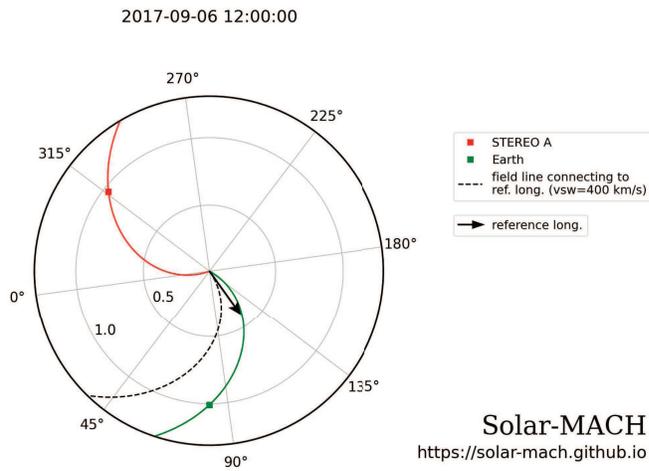}
\caption{Positions of the Earth (green) and STEREO-A (red) as seen
from the northern ecliptic pole on 6 September 2017. Also shown are
the Parker spirals connecting to Earth and STEREO-A, as well as the
reference longitude of the eruption given by the position of the
solar flare.} \label{fig:geometry}
\end{figure}

The solar eruption of 6 September 2017 was
magnetically well connected to Earth, as shown in
Figure~\ref{fig:geometry}, which depicts the Parker spirals
connecting to Earth (green) and STEREO-A (red) for a nominal solar
wind speed of 400~km\,s$^{-1}$. The black arrow indicates the
direction of the eruption based on the location of the associated
flare, assuming radial propagation. Indeed, a solar energetic
particle (SEP) event was detected near Earth, but not at STEREO-A,
which was separated from the eruption by almost 180$^{\circ}$.

\begin{figure}[ht]
\centering
\includegraphics[width=0.6\linewidth]{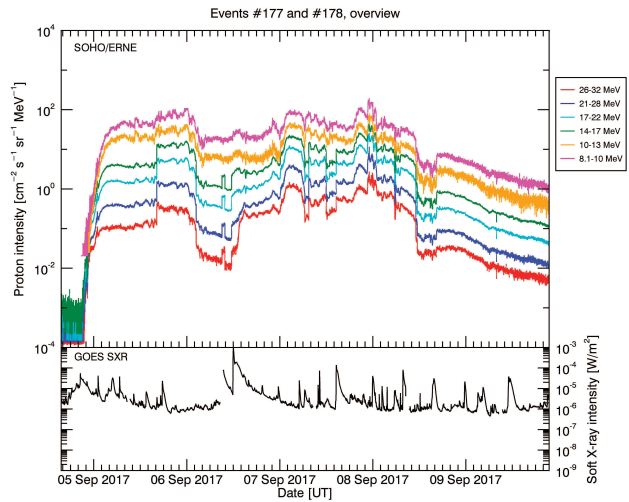}
\caption{Proton intensities as measured by the ERNE instrument
aboard SOHO in the period of 5 to 10 September 2017. For comparison,
the GOES SXR flux is shown in the lower panel. The increase of the
proton flux associated with the X9.3 flare peaking at 12:02~UT on 6
September is most clearly seen at the highest energies.
\label{fig:erne_ts}}
\end{figure}

Inspection of all available particle data near Earth has revealed
that at lower energies, the event is strongly contaminated by
particles associated with several previous events on September 5 and
6. The only clear signature of the event suitable for further
analysis was found in high-energy proton data as detected by the
ERNE instrument \citep{Torsti95} aboard SOHO. Figure~\ref{fig:erne_ts}
shows the ERNE proton intensities and the GOES soft X-ray flux in
the period of 5 to 10 September 2017. It is evident that our event
is best seen at the highest energies.

\begin{figure}[ht]
\centering
\includegraphics[width=0.6\linewidth]{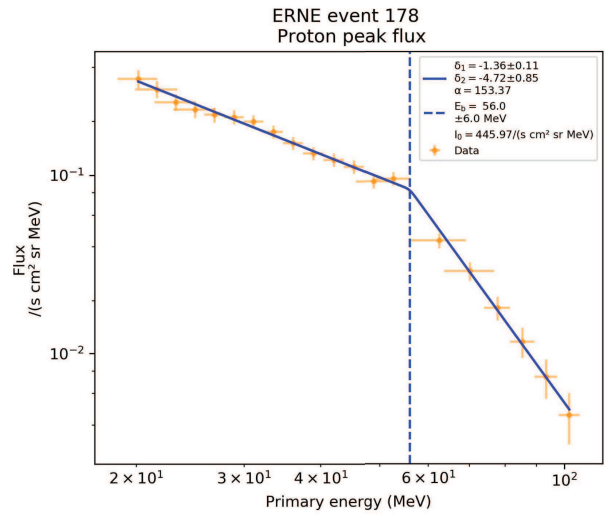}
\caption{Proton peak flux spectrum as derived from the ERNE data for
particle event of 6 September 2017 in the range of 20~MeV to
100~MeV. The data is shown as orange crosses, a spectral fit is
indicated as a blue line, and the fit parameters are indicated in
the legend. \label{fig:erne_spec}}
\end{figure}

In the following, we first consider the energetics of the prompt SEP
component, before deriving the energy of the total SEP population.
For the prompt component, we have used the ERNE data above 20~MeV to
obtain the proton peak flux spectrum which is shown in
Figure~\ref{fig:erne_spec}. The flux profiles were fitted with a
function, i.e., a modified Weibull profile given by \cite{Kahler17}.
We have fitted the peak flux spectrum $I_{\rm p}(E)$ with a broken
power-law \citep[cf.][]{Strauss2020}, which includes a smooth
transition at the spectral break according to:

\begin{equation}
I_{\rm p}(E) = I_0 \left( \frac{E}{E_0} \right) ^{\delta_1}  \left(
\frac{E^\alpha + E^\alpha_\mathrm{b}}{E^\alpha_0 +
E^\alpha_\mathrm{b}} \right) ^\frac{\delta_2-\delta_1}{\alpha} ,
\end{equation}
where E is particle energy, $I_0$ the flux at a reference energy
$E_0=0.1$~MeV, $\delta_1$ and $\delta_2$ the spectral indices below
and above the break energy $E_\mathrm{b}$, and $\alpha$ the
parameter describing the smoothness of the transition. The best-fit
parameters are indicated in Figure~\ref{fig:erne_spec}.

This SEP event was characterized by strong scattering, with
particles repeatedly passing back and forth through the position of
the spacecraft. We therefore adopt an approach
\citep[cf.][]{Wibberenz89} in order to relate the measured peak flux
spectrum $I_{\rm p}(E)$ at 1~AU to the proton fluence spectrum
injected at the source surface, which is based on diffusive
transport of protons in the interplanetary medium. Therefore, we can
use the peak flux spectrum as a measurement of the accelerated
proton spectrum at the Sun,
\begin{equation}
\frac{{\rm d}N}{{\rm d}E}=\frac{4 \pi r^3_0}{0.925 v} I_{\rm p}(E),
\end{equation}
where $N$ is the number of accelerated protons, $v$ is the particle
speed (recall that ${\rm d}E=v\,{\rm d}p$), and $r_0$ is the
distance from the Sun. ${\rm d}N/{\rm d}E$ represents the total
number of accelerated particles per steradian at the source surface
and per unit energy.

For the prompt SEP component, we have to consider two different
scenarios for constraining the area on the source surface over which
energetic particles are injected: either acceleration at a
CME-driven shock wave, or acceleration (and subsequent escape along
open magnetic field lines) in the solar flare. Considering the first
possibility and assessing the extent of the CME (cf.
Figure~\ref{en_cme}), adopting a solid angle of 2$\pi$ (i.e., a full
hemisphere) appears to be reasonable. Integrating from 20~MeV up to
100~MeV and multiplying with particle energy, we finally obtain an
energy in accelerated protons above 20~MeV of $1.4 \times
10^{28}$~erg. Assuming that the power-law extends to lower energies
with the same index $\delta_1$, we can derive estimates for the
total proton energy content. Extrapolating down to 50~keV, we get an
energy of $5.1 \times 10^{28}$~erg.

Next, we consider the alternative scenario where the protons are
accelerated in the solar flare and subsequently propagate to
interplanetary space along open magnetic field lines. We used a PFSS
(potential field source surface) extrapolation based on a synoptic
SDO/HMI map to determine the area on the source surface that
connects back to the flaring area. For the source surface, we used
three different heights: 2.5, 2.0 and 1.5 solar radii (heliocentric
distance). The standard height is usually taken as 2.5 solar radii,
but \cite{Virtanen20} have shown that the source surface has, for
the last two decades, seemed to stay below 2.0 solar radii.
Surprisingly, we found that there are no open magnetic field lines
connecting to the flare site, even if we lower the source surface
all the way down to 1.5 solar radii. While active regions are
predominantly closed-field regions, normally a PFSS extrapolation
does show some open field lines. We cannot rule out the possibility
of some open field lines were formed by magnetic reconnection in the
flare site, since the synoptic HMI maps and the PFSS model cannot
take into account those relatively short-lived processes. However,
even if this actually had been the case, the corresponding area on
the source surface would have been small. Taking the flare area as
an estimate, we derive a total SEP energy of only $5.1 \times
10^{28}$~erg.

\begin{figure}[ht]
\centering
\includegraphics[width=0.6\linewidth]{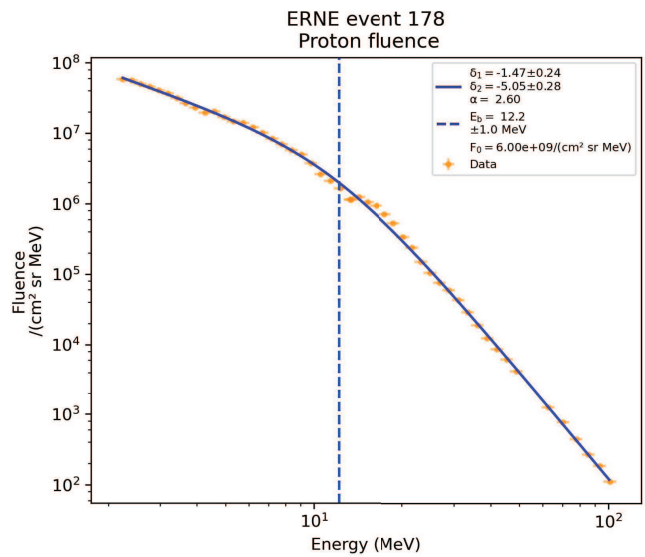}
\caption{Proton fluence spectrum as derived from the ERNE data for
particle event of 6 September 2017 in the range of 2~MeV to 100~MeV.
The data is shown as orange crosses, a broken power-law fit is
indicated as a blue line, and the fit parameters are indicated.
\label{fig:erne_spec_fluence}}
\end{figure}

We now consider the energetics of the total SEP population. The work
by \cite{Mewaldt08} is widely recognized as the pioneering in SEP
energetics, and a comprehensive study has been given by
\cite{Emslie12}. For a sample of 38 solar eruptive events, they
obtained SEP energetics ranging from $1.3 \times 10^{29}$~erg to
$4.3 \times 10^{31}$~erg. In order to compare with these results, we
apply their methodology to our event. The observed fluence (in units
of cm$^{-2}$) at 1~AU is converted to on-axis fluence assuming a
Gaussian in latitude and longitude with a width of $\sigma =
38^\circ$ centered at the flare location (9S,34W) and a solar wind
speed of 450~km\,s$^{-1}$, i.e., a connection point at (7N,52W).
This implies an increase of the fluence by a factor of 1.22. Then
the Gaussian fluence distribution is integrated over longitude and
latitude and multiplied by (1 AU)$^2$; this implies a further factor
of $6.21 \times 10^{26}$~cm$^2$. Finally, we apply a factor 0.5 to
account for the multiple-crossing effect. Thus, the total factor to
apply here is $3.8 \times 10^{26}$~cm$^{2}$. We obtain the total
proton fluence from the ERNE spectrum shown in
Figure~\ref{fig:erne_spec_fluence} by integrating the spectrum
50~keV to 400~MeV and applying the correction factor. This gives a
total energy of about $10^{31}$ erg, which is already comparable to
the largest values derived by the previous finding
\citep[cf.][]{Emslie12}. Thus, the prompt component of this SEP
event contains a negligible fraction of the SEP energy, and almost
all of it is due to the interplanetary acceleration.

\section{Discussions}
The X9.3 flare on 6 September 2017 is the most powerful flare during
the solar cycle 24, and it has been studied by several authors
\citep[e.g.,][]{Dominique18,Kolotkov18,Lysenko19,Karlicky20,Motorina20,Lee21,Li21,Zhang21b}.
In this work, we explored the energy partition in the X9.3 flare.
Based on multi-instrument measurements, we calculated the flare
radiated energy in X-ray and UV wavelength ranges, as listed in
Table~\ref{tab_rad}. It can be seen that the maximal energy during
the X9.3 flare is measured by LYRA at 0.1$-$80~nm, which is about
3.5$\times$10$^{31}$~erg. It is roughly equal to that measured by
EVE/ESP~0.1$-$7~nm, such as 3.4$\times$10$^{31}$~erg. They are
$\sim$20 times bigger than the radiated energy in GOES~0.1$-$0.8~nm,
which is similar to previous observations \citep{Zhang19,Cai21}. The
flare radiated energy observed by LYRA in 0.1$-$20~nm is about
3.1$\times$10$^{31}$~erg, and it is a bit smaller than that in
ESP~0.1$-$7~nm. This is mainly because that there is a drop of the
spectral response in the middle of the bandpass in this channel, for
instance, a gap in the responsivity between 2$-$6~nm
\citep[cf.][]{Dolla12,Dominique13}. On the other hand, the flare
energy radiated in EUV wavelengths is in the order of 10$^{29}$~erg,
which is much smaller than that radiated in the SXR channel, i.e.,
two orders of magnitude. While the flare radiated energy in
Ly$\alpha$ and MUV increases slightly, could be about 10$^{30}$~erg,
but still less than that in the SXR channel of 0.1$-$7~nm. Our
observations suggest that the flare emission mainly comes from the
shorter wavelength, for instance, the major contribution of the
flare radiation comes from the SXR emission at 0.1$-$7~nm, while
there is actually quite few contribution coming from the EUV ranges.
It should be pointed out that here the flare radiation between
0.1$-$222~nm is studied due to observational limitations, and thus
the observational result is only applicable to this wavelength
range. In fact, most of the flare energy is radiated in the
wavelength longward of 200~nm, for instance, the radiated energy in
the WL continuum could be as high as 70\%
\citep[cf.][]{Kretzschmar11,Kleint16}. However, the major
contribution from the WL continuum is impossible estimated due to
the lack of observations, since the HMI continuum represents the
true WL continuum only in the quiet sun, while it can be significant
deviations in the flare \citep{Svanda18}. On the other hand, a
majority of the flare energy is released in the wavelength shortward
of 27~nm, i.e., about 19\% , while very little flare energy is
radiated in EUV passbands \citep[e.g.,][]{Woods04,Kretzschmar10}.
Therefore, we could assume that the total solar irradiance (TSI) of
the X9.3 flare is about 3$\times$10$^{32}$~erg. We wanted to state
that the flare energy in the radio emission is not estimated, mainly
due to its rather low contribution to the total TSI
\citep{Pasachoff78}.

According to the standard flare model
\citep{Masuda94,Priest02,Shibata11}, the flare radiated energy in
multiple wavelengths could convert from either thermal or nonthermal
energies. In this study, we also calculate the peak thermal energy
of SXR-emitting plasmas and the nonthermal energy, which are
estimated to about 3.1$\times$10$^{31}$~erg and
4.5$\times$10$^{31}$~erg. Similar to previous observations
\citep{Emslie12,Feng13,Zhang19,Cai21}, they are adequate sufficient
to support the radiative loss energy (U$_{rad}$) of hot plasmas, as
seen in Table~\ref{tab_flare}. However, the thermal and nonthermal
energies are smaller than the TSI of the X9.3 flare in full
wavebands. The peak thermal energy from GOES is based on the
assumption of an isothermal temperature \citep{White05,Emslie12},
but the solar flare often shows multi-thermal temperatures, for
instance, the multi-thermal flare loops \citep{Aschwanden15,Li21r}.
Hence, our measurement of the peak thermal energy could be regarded
as the lower limit estimation using two GOES SXR fluxes. The
radiative loss energy in our case is a little smaller than the
previous result \citep{Motorina20}, who used a much long-time
integration between 11:53$-$13:40~UT. We use the short-time
integration because that the GOES flux (Figure~\ref{img}~a) shows a
growth trend after $\sim$12:30~UT, which might be regarded as
another flare. Moreover, the EUV, Ly$\alpha$, and MUV fluxes do not
exhibit the similar growing trend. We should state that the energy
spectrum of the X9.3 flare captured by \textit{Insight}-HXMT is from
about 100~keV to 800~keV \citep{Zhang21b}, and here we only fit the
energy range between 200$-$600~keV every 10~s to improve the signal
to noise ratio. Our estimation of the nonthermal energy is larger
than the total ion energy (1.1$\times$10$^{31}$~erg) estimated from
the Konus-Wind observation \citep{Lysenko19,Motorina20}. This is
because that the available observational data of Konus-Wind only
remains about 72~s, which is much shorter than \textit{Insight}-HXMT
observations (Figure~\ref{img}~b). Similar to Konus-Wind
\citep{Lysenko19}, \textit{Insight}-HXMT misses the credible data
below the energy of 200~keV for the X9.3 flare. Thus, the energy
spectrum fitting tend to the higher energetic electrons/pronotos,
but lack of the lower energetic electrons. Hence, our measurement of
the nonthermal energy is also the minimum estimation using the
\textit{Insight}-HXMT spectrum. Therefore, the actual thermal and
nonthermal energies could be large enough to support the TSI of the
X9.3 flare.

Based on WL continuum images measured by LASCO/C2 and C3, we
estimate the mass and height of the CME followed by the X9.3 flare,
as well as the propagation speed of the CME. Then, we obtain its
kinetic and gravitational potential energies, as shown in
Figure~\ref{en_cme}. It can be seen that the kinetic energy of the
CME is a bit larger than the gravitational potential energy, which
is mainly related to the fast propagation speed of the CME
\citep{Aschwanden16}. Our observational results agrees with previous
findings for the energy distribution of CMEs
\citep{Emslie04,Emslie12}. Conversely, some authors
\citep{Vourlidas00,Ying18} also found that the majority CME energy
is from the gravitational potential energy, but for those CMEs have
slower propagation speeds, i.e., $<$500~km~s$^{-1}$. The total CME
energy here is estimated to $\sim$4$\times$10$^{32}$~erg, which is
roughly equal to the flare energy \citep{Emslie05,Feng13,Ying18}. We
want to stress that the magnetic energy taken by the CME is not
estimated, since the magnetic field strength in the CME is unknown.

It is well accepted that both the flare and CME energies are
released from the magnetic fields in solar disk, such as the AR.
Hence, we calculate the magnetic free energy released from the AR
NOAA 12673 on the Sun. Using the SDO/HMI vector magnetogram before
the X9.3 flare, the nonpotential and potential fields are estimated
by the NLFFF extrapolation
\citep{Wheatland00,Wiegelmann06,Wiegelmann12}. Then, the magnetic
free energy can be estimated to roughly 2$\times$10$^{33}$~erg,
which is one order of magnitude larger than the flare and CME
energies. Our estimations suggest that the magnetic free energy
released by the solar AR is adequate enough to power the major flare
and its accompanied CME, which agrees with previous findings for the
eruptive \citep{Feng13,Aschwanden16}.

At last, we have estimated the energy content of the SEPs. The
contribution from prompt SEPs was negligible, which is consistent
with the strong magnetic containment in the associated flare.
Comparing the total SEP energy with the CME energy, we find that the
SEPs contain about 2.5\% of the CME energy. This is consistent with
previous values obtained by \cite{Emslie12} who
found fractions of 1--10\%, and shows that the CME was quite
efficient in accelerating SEPs in interplanetary space.

\section{Summary}
Using coordinated observations measured by multiple instruments, for
instance, GOES, \textit{Insight}-HXMT, PROBA2/LYRA, SDO/EVE, SDO/HMI,
SDO/AIA, Hinode/XRT, SOHO/LASCO, and SOHO/ERNE, we evaluate the
energetics of a powerful flare, the accompanied CME and SEPs on 6
September 2017. Our major results are summarized in following:

\begin{enumerate}

\item We focus on the flare radiated energy in SXR, EUV, Ly$\alpha$, and
MUV wavelength ranges, although the major contribution of the TSI is
from the WL continuum. However, we did not find the WL continuum
data for the X9.3 flare. In the studied wavelength ranges, the flare
radiated energy in the SXR~0.1$-$7~nm is much larger than other
wavebands, which is $\sim$3.4$\times$10$^{31}$~erg. Only a small
amount of flare energy emitted at the EUV wavelength, such as in the
order of 10$^{29}$~erg. At wavebands of Ly$\alpha$ and MUV, the
flare radiated energy could be in the order of 10$^{30}$~erg.

\item The total radiated energy of the X9.3 flare could be as high as
$\sim$3$\times$10$^{32}$~erg, which is almost in the energy realm of
the stellar flare. The peak thermal energy is estimated to be
$\sim$3.1$\times$10$^{31}$~erg, assuming an isothermal temperature
for the flare. The nonthermal energy is also estimated to
$\sim$4.5$\times$10$^{31}$~erg. Both the thermal and nonthermal
energies are minimum estimations, mainly due to the limitation of
observational data.

\item The energy content of the accompanied CME is estimated, which has a
total energy of $\sim$4$\times$$10^{32}$~erg. The kinetic energy of
the CME is larger than the potential energy, and they both increase
with the altitude.

\item The released energy from the X9.3 flare is comparable to the energy
carried by the CME. The magnetic free energy stored in the AR is
estimated to about 2$\times$$10^{33}$~erg with the NLFFF
extrapolation, and it is able to support the powerful flare and CME
eruptions.

\item The SEP energy content is estimated as about 10$^{31}$~erg for
protons in the range of 50~keV to 400~MeV, which amounts to 2.5\% of
the CME energy. A negligible fraction was contributed by the prompt
SEP component. Thus, the CME was very efficient in accelerating
particles in interplanetary space.

\end{enumerate}

\normalem
\begin{acknowledgements}
The authors would like to thank the referee for his/her valuable
comments and suggestions. We thank the teams of GOES,
\textit{Insight}-HXMT, LYRA, SDO, Hinode, and SOHO for their open
data use policy. This work is funded by the National Key R\&D
Program of China 2022YFF0503002 (2022YFF0503000), the NSFC under
grant 11973092, 12073081, 12003064, 12103090, U1938102, the
Strategic Priority Research Program on Space Science, CAS, Grant No.
XDA15052200 and XDA15320301. This project is also supported by the
Specialized Research Fund for State Key Laboratories. LYRA is a
project of the Centre Spatial de Liege, the
Physikalisch-Meteorologisches Observatorium Davos and the Royal
Observatory of Belgium funded by the Belgian Federal Science Policy
Office (BELSPO) and by the Swiss Bundesamt f\"{u}r Bildung und
Wissenschaft. Part of this work was performed in the framework of
the SERPENTINE project, which has received funding from the European
Union's Horizon 2020 research and innovation program under grant
agreement No. 101004159. N. D. is grateful for support by the Turku
Collegium for Science, Medicine and Technology of the University of
Turku, Finland. M. D. acknowledges support from the Belgian Federal
Science Policy Office (BELSPO) in the framework of the ESA-PRODEX
program, grant No. 4000134474.
\end{acknowledgements}

\label{lastpage}
\clearpage

\end{document}